\documentclass[a4paper]{jpconf}

\usepackage{graphicx}

\newcommand{\be}{\begin{equation}}
\newcommand{\ee}{\end{equation}}
\newcommand{\beqn}{\begin{eqnarray}}
\newcommand{\eeqn}{\end{eqnarray}}
\newcommand{\dd}{{\mathrm d}}
\newcommand{\Dirac}{\rlap {\hspace{-0.01mm} \slash} D}
\newcommand{\dirac}{\rlap {\hspace{-0.5mm} \slash} \partial}

\begin{document}
\title{Influence of quark masses on the QCD phase diagram in the presence of a magnetic field}
\author{A.~J.~Mizher}
\address{Instituto de Ciencias Nucleares, Universidad Nacional Aut\' onoma de M\' exico, Apartado Postal 70-543, M\' exico Distrito Federal 04510, M\' exico}
\ead{ana.silveira@correo.nucleares.unam.mx}

\begin{abstract}

The most recent lattice results on the QCD phase diagram in the presence of a strong magnetic field strongly disagree with results from previous lattice simulations and several model calculations. The most remarkable difference is the qualitative behavior of the critical temperature as a function of the magnetic field intensity, which in later results are shown to decrease in opposition to what was found previously. According to the authors, such a discrepancy in lattice simulations could be due to different lattice spacing, or different number of flavors and quark masses. We investigate the influence of quark masses on the Polyakov-Quark-Meson model and show that, although quantitatively the results are sensitive to those parameters, the qualitative behavior remains the same.

\end{abstract}

\vspace{0.5cm}

{\Large{\bf Introduction}}

\vspace{0.5cm}

The phase structure of the QCD phase diagram in the presence of magnetic field was studied in recent years by several model calculations and also by full lattice QCD simulations. Such an interest on the subject was motivated by the fact that it was proposed that a very intense magnetic field, of order $5-10m_\pi^2$ \cite{Kharzeev:2007jp,Skokov:2009qp}, must be produced in non-central heavy-ion collisions and must have a considerable influence in the phase transitions suffered by the matter generated after the collision. 

Most of the model calculations \cite{Johnson:2008vna,Mizher:2010zb,Gatto:2010qs,Gatto:2010pt} indicate an increasing behavior of the critical temperature as a function of the magnetic field and early lattice simulations \cite{D'Elia:2010nq} corroborated these results. Later lattice results \cite{Bali:2011qj}, however, indicate a decrease of the critical temperature with the magnetic field intensity. In order to check their code and algorithm simulation, the authors of the latter have reproduced the results of the former group applying the same setup. Since in this context the previous results were recovered, the discrepancy was then associated to different lattice spacing or difference in number of flavors and quark masses. Approaches beyond mean field were considered \cite{Skokov:2011ib}, still obtaining an increase of the critical temperature with the magnetic field. A later approach  \cite{Fraga:2012fs} could reproduce qualitatively the lattice results in \cite{Bali:2011qj}, considering the context of the MIT bag model.

According to \cite{Bali:2011qj} the difference between their result and the previous one can be traced back to the behavior of the chiral condensate as a function of the magnetic field. While in \cite{D'Elia:2010nq} it was found that the condensate has a monotonically increasing dependence on the field for all the range of temperature tested, \cite{Bali:2011qj} obtains a more complex behavior, in which such a dependence can increase or decrease depending on the temperature. In order to get some insight on the role of quark masses in model calculations we use the linear sigma model coupled to quarks and to the Polyakov loop to obtain a phase diagram as a function of the magnetic field for different values of constituent quark masses. Although we include the Polyakov loop, taking into account the interplay between the chiral and deconfining transition, we limit our analysis to the chiral condensate, following the indication in \cite{Bali:2011qj} that the chiral transition guides the general behavior of both transitions. 

\vspace{0.5cm}

{\Large{\bf The Polyakov-Quark-Meson model}}

\vspace{0.5cm}

In this context, the deconfinement properties are encoded in the expectation value of the Polyakov loop, which works as an order parameter for this transition:
\beqn
\mbox{Confinement}:\quad
\left\{
\begin{array}{llll}
\langle L \rangle  & = & 0 \quad , \quad & \mbox{low $T$}   \\
\langle L \rangle  & \neq & 0 \quad , \quad & \mbox{high $T$}
\end{array}
\right.\,, \qquad\quad L(x) = \frac{1}{3} \Tr {\cal P} \exp \Bigl[i
\int\limits_0^{1/T} \dd \tau \, A_4(\vec x, \tau) \Bigr]\,,
\label{eq:L} \label{eq:L:phases} \eeqn
where $A_4$ is the matrix-valued temporal component of the Euclidean gauge field $A_\mu$ and the symbol ${\cal P}$ denotes path ordering. On the other hand, the expectation value of the chiral condensate works as the order parameter for the chiral transition, indicating if this symmetry is broken or restored for each value of external parameters:

\beqn
\mbox{Chiral symmetry}:\quad
\left\{
\begin{array}{llll}
\langle \sigma \rangle  & \neq & 0 \quad , \quad & \mbox{low $T$}   \\
\langle \sigma \rangle  & = & 0 \quad ,\quad & \mbox{high $T$}
\end{array}
\right.\,,
\label{eq:sigma:phases}
\qquad\qquad
\begin{array}{lll}
\phi & = & (\sigma,\vec{\pi})\,,\\
\vec{\pi} & = & (\pi^{+},\pi^{0},\pi^{-})\,.
\end{array}
\label{eq:phi}
\eeqn
The chiral sector coupled to quarks is given by the Lagrangian:
\begin{eqnarray}
{\cal L} &=&
 \overline{\psi}_f \left[i\gamma ^{\mu}\partial _{\mu} - g(\sigma +i\gamma _{5}
 \vec{\tau} \cdot \vec{\pi} )\right]\psi_f 
+ \frac{1}{2}(\partial _{\mu}\sigma \partial ^{\mu}\sigma + \partial _{\mu}
\vec{\pi} \partial ^{\mu}\vec{\pi} )
- V(\sigma ,\vec{\pi})\;,
\label{lagrangian}
\end{eqnarray}
which contains spontaneous and explicit symmetry breaking and the coupled of mesons and quarks via a Yukawa term. The quarks act as a heat bath in which the condensate evolves and quark loop corrections provides the thermal contribution to the potential. The implementation of the magnetic field is done replacing the simple derivative that describes the dynamics of the fermions by covariant derivative. Such a derivative contains an Abelian gauge field which can be seen as a vector potential for the magnetic field. Also the interaction with the pure gauge sector, through the Polyakov loop, can be implement in the same fashion, the Polyakov loop being parametrized in such a way that a non-Abelian gauge field interacting with the mesons can represent it. The 1-loop quark correction is then given by:
\beqn
e^{i V_{3d} \, \Omega_q/T}  =
\left[\frac{\det(i \Dirac^{(q)} - m_q )}{\det (i \dirac - m_q)} \right] 
 \cdot
\left[\frac{\det_T(i \Dirac^{(q)} - m_q)}{\det (i \Dirac^{(q)} - m_q )}\right]\,,
\label{eq:Omega:quark2}
\eeqn
where the quark mass is given by $m_q=\sqrt{m_c^2 + (g\langle \sigma \rangle)^2}$, with $m_c$ the constituent mass, and the covariant derivative contains both gauge fields, the Abelian and the non-Abelian. 

To reproduce the nature of the phase transition predicted in \cite{Bali:2011qj}, a crossover, we take into account the quark degrees of freedom in the vacuum. The non inclusion of this correction implies a first order phase transition and a discussion regarding this point is done in \cite{Mizher:2010zb}. 

We consider several values for the constituent quark mass in a range between the chiral limit and the physical consituent quark mass and also an extrapolation of that, $m_q=500$ MeV, in order to investigate the systematics. For a set of temperature values, we obtain the condensate as a function of the magnetic field. The results are plot in Fig.1 
. In the chiral limit and for small values of the quark mass the condensate increases with the magnetic field and saturates at a constant value for all values of the temperature. This behavior is again in accordance with \cite{D'Elia:2010nq}, which shows in their Fig.3 a similar saturation (regarding that the periodic behavior is an artefact from the fact that they have introduced the $U(1)$ field as a phase factor $A e^{i\phi}$). As the quark mass is increased the value of magnetic field to reach saturation becomes larger, however the behavior is systematically the same. 

\begin{figure}[!thb]
\begin{center}
\begin{tabular}{cccc}
\includegraphics[width=70mm,clip=true]{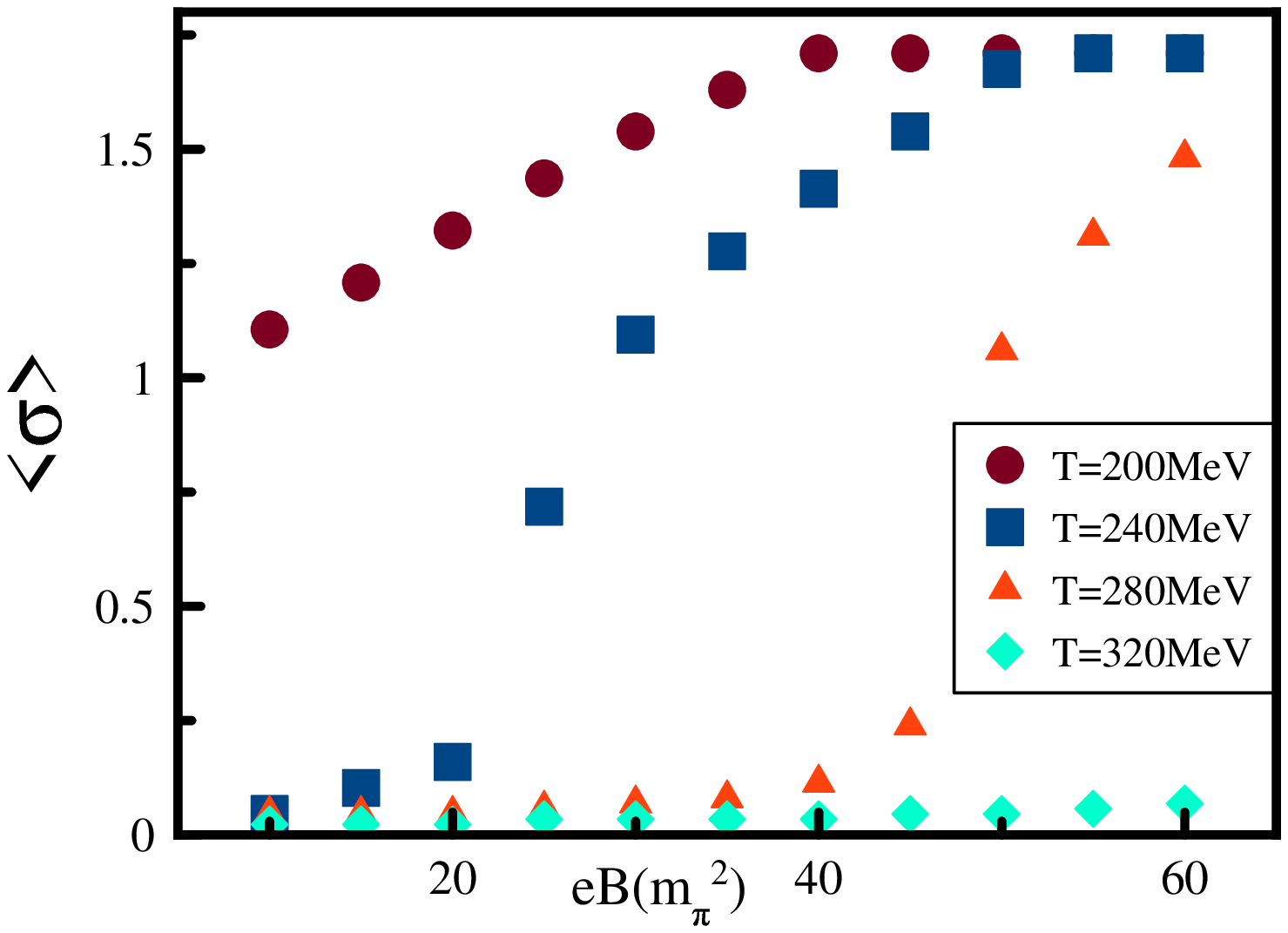} 
& \includegraphics[width=70mm,clip=true]{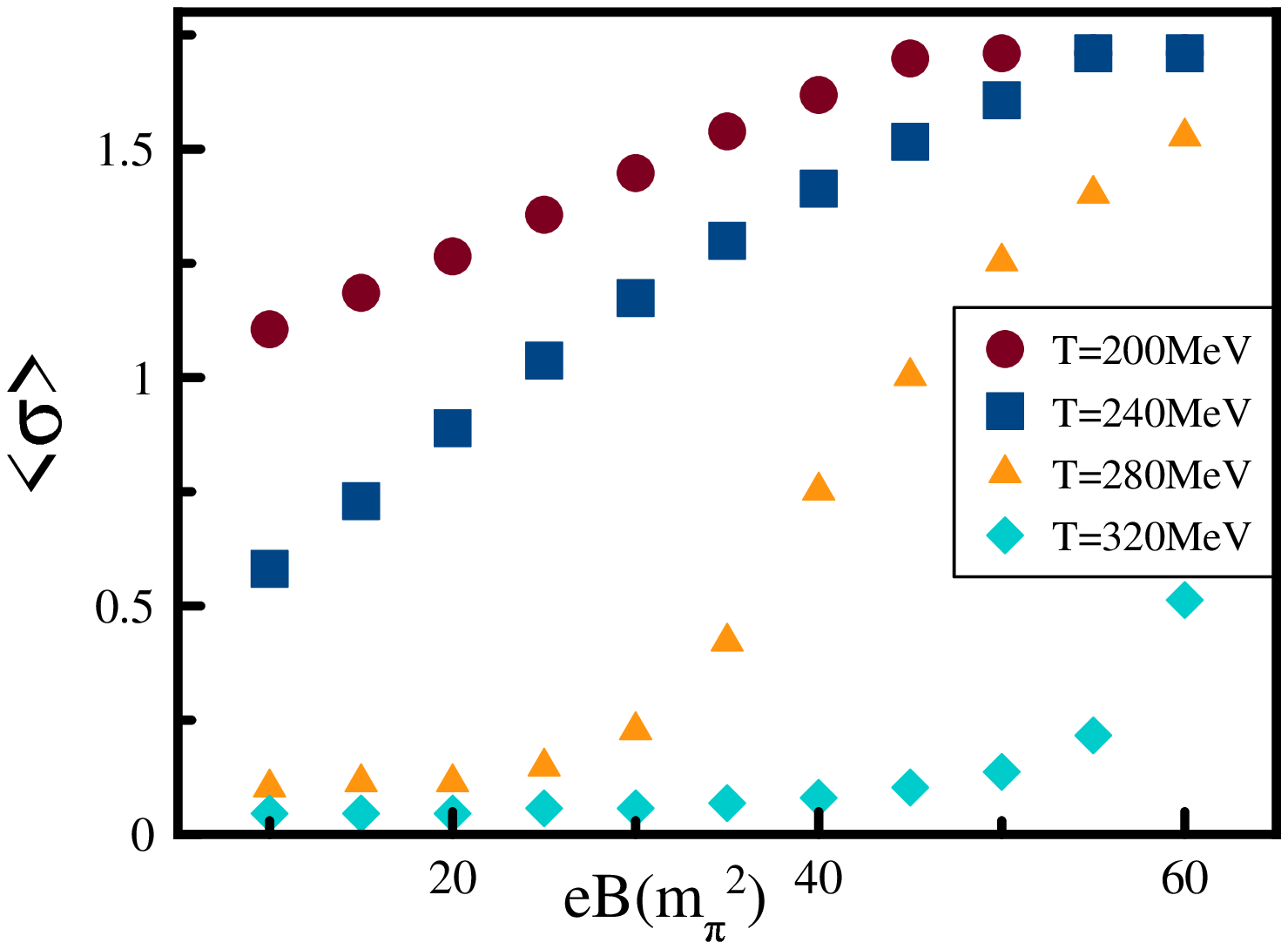}\\[-3mm]\\
 (a) & (b) \\
\end{tabular}
\end{center}
\label{fig:fixT}
\end{figure}
\begin{figure}[!thb]
\begin{center}
\begin{tabular}{cccc}
\includegraphics[width=70mm,clip=true]{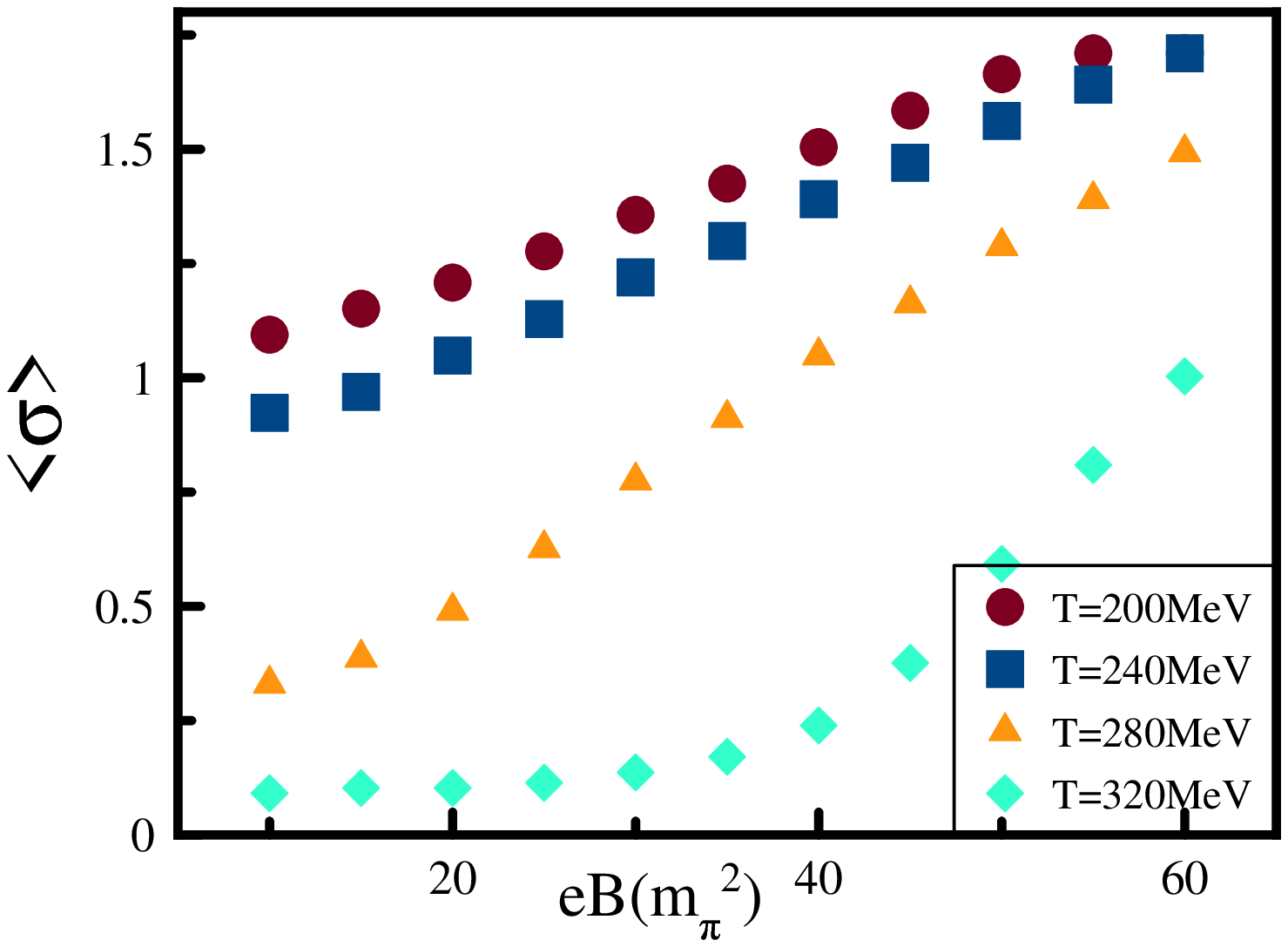} \\
(c)\\
\end{tabular}
\end{center}
\caption{Expectation value of the chiral condensate for different values of magnetic fields. The different plots differ by the value of the quark masses considered and the different curves by temperature. We considered (a) $m_q=0$ MeV, (b) $m_q=300$ MeV and (c) $m_q=500$ MeV.}
\label{fig:ximin_B}
\end{figure}

\vspace{0.5cm}

{\Large{\bf Conclusions}}

\vspace{0.5cm}

In this work we have studied the role of quark masses for the chiral phase transition in the context of the Polyakov-Quark-Meson model in the presence of a magnetic field. Motivated by discrepant lattice results regarding the behavior of the chiral condensate as a function of the magnitude of the magnetic field we have studied the effect of going beyond the chiral limit, since this was presented as a potential cause for the different lattice results. We showed that for all values of quark masses considered, from chiral limit to physical constituent quark masses, the qualitative behavior doesn't change, in accordance with \cite{D'Elia:2010nq} and in disagreement with more refined results in \cite{Bali:2011qj}. A systematic saturation of the value of the chiral condensate for all values of temperature is found, also in agreement with \cite{D'Elia:2010nq}. As a consequence, the critical temperature increases with the magnetic field and it was seen no sign in favor of a decreasing for any value of the magnetic field. 

\vspace{0.5cm}

{\Large{\bf Acknowledgments}}

\vspace{0.5cm}

A.J.M. thanks Cristian Villavicencio for discussions. Support for this work has been received in part from DGAPA-UNAM under Grant No. PAPIIT-IN103811 and CONACyT-M\'exico under Grant No. 128534.


\vspace{1cm}

{\Large{\bf Bibliography}}

\end{document}